\documentclass[12pt]{article}
\usepackage{epsf}
\usepackage{cite}
\begin{document}
\begin{center}
{\Large\bfseries
Mass Formulae for Supernarrow Dibaryons and Exotic Baryons}
\vskip 5mm

L.V. Fil'kov
\vskip 5mm

{{\small\it Lebedev Physical Institute, Moscow, Russia}

{\it E-mail: filkov@sci.lebedev.ru}}
\end{center}

\vskip 5mm

\begin{abstract}

The mass formulae for the supernarrow dibaryons and the exotic baryons
with small masses are constructed. With this aim their self energies
are calculated in one loop approximation using the dispersion relations
with two subtractions. The values of the masses obtained in this approach
are in a good agreement with the experimental data. The mass formula for
the baryons is also used to calculate the mass of the $\Delta(1232)$
resonance.
\end{abstract}

\vskip 10mm

\section{Introduction}

Supernarrow dibaryons (SNDs) are 6-quark states,
a decay of which into two nucleons is forbidden by the Pauli exclusion
principle \cite{fil1,fil2,alek,alek1}.
Such dibaryons satisfy the following condition:
\begin{equation}
(-1)^{T+S}P=+1
\end{equation}
where $T$ is the isospin, $S$ is the internal spin, and $P$ is the
dibaryon parity.
These dibaryons with the masses \mbox{$M < 2m_{N}+m_{\pi}$}
($m_N (m_{\pi}$) is the nucleon (pion) mass) can mainly decay
by emitting a photon. This is a new class of dibaryons
with the decay widths \mbox{$< 1$keV}.
The experimental discovery of such states would have important
consequences for particle and nuclear physics and astrophysics
\cite{kolom}.

The SNDs have been first experimentally observed in works \cite{konob,izv,
yad,prc,can,dub,epj} with the masses $M=1904$, 1926, and 1942 MeV.
Unfortunately, existed theoretical models do not allow a correct
calculation of dibaryon masses in the mass region under consideration.

On the other hand, in the reactions
$pd\to p+pX_1$ \cite{can,dub,epj} and $pp\to \pi^+pX$ \cite{tat},
connected with the SND production, peaks have
been found in the missing mass spectra. These peaks have been interpreted
in \cite{tat} as a manifestation of new exotic baryons with anomaly small
masses. The calculation of the masses of exotic baryons are
usually realized
in the framework of the quark cluster models \cite{muld,besl,ueh,kon}.
However, these models do not reproduce all baryon correctly enough.
And although the question about the nature of these peaks is still open,
it is important to construct a mass formula which allows a more correct
analysis of them.

In the present paper we construct mass formulae for SNDs and exotic
baryons. With this aim we calculate self energies of the SNDs and
the exotic baryons in one loop approximation using dispersion
relations with two subtractions.

In order to check an efficiency of this approach, the mass formula for
the exotic baryons is also used to calculate the mass of the
$\Delta(1232)$ resonance.

The application of the mass formulae constructed for a calculation of
the masses of the SNDs and the exotic baryons results in a good agreement
with the experimental data \cite{epj,tat}.

The contents of the paper are the following. In Section 2 and 3 the works
on the SNDs and the exotic baryons are reviewed, respectively.
In Section 4 the mass formula for exotic baryons is constructed and the
equation obtained is used for the calculation of the mass of the
$\Delta(1232)$ resonance. The calculation of the masses of the exotic
baryons is performed in Section 5. The construction of the mass formula
for SNDs and the calculation their masses are given in Section 6. The main
conclusion are given in Section 7.

\section{Supernarrow dibaryons}

In works \cite{konob,izv,yad,prc,can,dub, epj}, the reactions
$pd\to p+pX_1$ and
$pd\to p+dX_2$ were studied with the aim of searching for SNDs.
The experiment was
carried out at the Proton Linear Accelerator of INR with 305 MeV
proton beam using
the two-arm spectrometer TAMS. As was shown in \cite{yad,prc},
the nucleons and the deuteron from the SND decay into $\gamma NN$
and $\gamma d$ have to be emitted in a narrow angle cone with respect to
the direction of the dibaryon motion. On the other hand, if a dibaryon
mainly decays into two nucleons, then the expected  angular cone of the
emitted nucleons must be more than $50^{\circ}$. Therefore, a detection
of the scattered proton in coincidence with the proton (or the deuteron)
from the decay of the dibaryon at correlated angles allowed the authors
to suppress the contribution of the background processes essentially and
to increase the relative contribution of a possible SND production.

Several software cuts have been applied to the mass spectra in these works.
In particular, the authors limited themselves by the consideration of an
interval of the proton energy from the decay of the $pX_1$ states, which
was determined by the kinematics of the SND decay into $\gamma NN$ channel.
Such a cut is very important as it provides an additional possibility to
suppress essentially the
contribution from the background reactions and random coincidences.

In \cite{can,dub,epj}, CD$_2$ and $^{12}$C were
used as targets. The scattered proton was detected in the left arm
of the spectrometer TAMS at the angle $\theta_L=70^{\circ}$. The second
charged particle (either $p$ or $d$) was detected in
the right arm by three telescopes located at $\theta_R=34^{\circ}$,
$36^{\circ}$, and $38^{\circ}$.

As a result,
three narrow peaks in the missing mass $M_{pX_1}$ spectra have been
observed at
$M_{pX_1}=1904\pm 2$, $1926\pm 2$, and $1942\pm 2$ MeV  with widths
equal to the experimental
resolution ($\sim 5$ MeV) and with numbers of standard deviations of
6.0, 7.0, and 6.3, respectively. It should be noted
that the dibaryon peaks at $M_{pX_1}=1904$ and 1926 MeV had
been observed earlier by same authors in \cite{konob,izv,yad,prc}
at somewhat different kinematical conditions. On the other hand,
no noticeable signal of dibaryons has been observed  in the missing
mass $M_{dX_2}$ spectra of the reaction $pd\to p+dX_2$.
The analysis of the angular distributions of the protons from the decay of
the $pX_1$ states and the suppression observed of the SND decay into
$\gamma d$ showed that the peaks found can be explained as a
manifestation of the isovector SNDs, the decay of which into two nucleons
is forbidden by the Pauli exclusion principle.

An additional information about the nature of the observed states
was obtained by studying the missing mass $M_{X_1}$ spectra of the
reaction $pd\to p+pX_1$.
If the state found is a dibaryon mainly decaying into two nucleons then
$X_1$ is a neutron and the mass $M_{X_1}$ is equal to the neutron mass
$m_n$. If the value of $M_{X_1}$, obtained from the experiment, differs
essentially from $m_n$ then $X_1=\gamma+n$ and it is an additional
indication that the dibaryon observed is the SND.

The simulation of the missing mass $M_{X_1}$ spectra of the reaction
$pd\to p+pX_1$ has been performed \cite{can,dub,epj}
assuming that the SND decays as
SND$\to \gamma+^{31}\!S_0\to \gamma pn$ through two-nucleon singlet state
$^{31}S_0$ \cite{fil2,prc,epj}. As a result, three narrow peaks at
$M_{X_1}=965$, 987, and 1003 MeV have been predicted. These peaks
correspond to the decay of the isovector SNDs with the masses 1904, 1926,
and 1942 MeV, respectively.

In the experimental missing mass $M_{X_1}$ spectrum
besides the peak at the neutron mass
caused by the process $pd\to p+pn$,
a resonance-like behavior of the spectrum has been observed at $966\pm 2$,
$986\pm 2$, and $1003\pm 2$ MeV \cite{can,dub,epj}.
These values of $M_{X_1}$ coincide with
the ones obtained  from the simulation and essentially differ from
the value of the neutron mass (939.6 MeV). Hence, for all states under
study, one has $X_1=\gamma+n$ in support of the statement that the
dibaryons found are SNDs.

In \cite{khr} dibaryons with exotic quantum numbers were searched for
in the process $pp\to pp\gamma\gamma$. The experiment was performed with
a proton beam from the JINR Phasotron at an energy of about 216 MeV. The
energy spectrum of the photons emitted at $90^{\circ}$ was measured.
As a result, two peaks have been observed in this spectrum. This behavior
of the photon energy spectrum was interpreted as a signature of the exotic
dibaryon resonance with the mass of about 1956 MeV and a possible isospin
$T=2$.

On the other hand, an analysis \cite{cal} of the Uppsala proton-proton
bremsstrahlung data looking for the presence of a dibaryon in the mass
range from 1900 to 1960 MeV gave only the upper limits of 10 and 3 nb for
the dibaryon production cross section at proton beam energies of 200 and
310 MeV, respectively.
This result agrees with the estimates of the cross
section obtained at the conditions of this experiment in the framework
of the model suggested in \cite{fil2,prc} and does
not contradict to the data of \cite{khr}.

It should be noted that if this dibaryon has $T=2$ then it could
not be a SND, the decay of which into two nucleons is forbidden by the
Pauli exclusion principle.
The decay of such a dibaryon into two nucleons is forbidden by
the value of the isospin $T=2$.

The reactions $pd\to pdX_2$ and $pd\to ppX_1$ have
been also investigated by Tamii {\em et al.} \cite{tamii}
at the Research Center for Nuclear Physics
at the proton energy 295 MeV in the mass region of 1896--1914
MeV. They did not observe any narrow structure in this mass region
and obtained the upper limit of the production cross section of a
NN-decoupled dibaryon was equal to \mbox{$\sim 2\,\mu$b/sr} if the dibaryon
decay width $\Gamma_D<< 1$ MeV. This limit is
smaller than the value of the cross section of \mbox{$8\pm 4\,\mu$b/sr}
declared in \cite{prc}.

However, the latter value was overestimated that was caused by
not taking into account
angle fluctuations related to a beam position displacement on
the CD$_2$ target during the run.
As was shown in the next experimental runs, the real
value of the cross sections of the production of the SND with
the mass 1904 MeV had to be smaller by 2--3 times than
that was estimated in \cite{prc}.

On the other hand, the simulation showed that the energy distribution
of the protons from the decay of the SND with the mass of 1904 MeV
has to be rather narrow with the maximum at $\sim 74$ MeV. This
distribution occupies the energy region of 60--90 MeV. The experiments
\cite{prc,epj} confirmed the result of this simulation. However,
in \cite{tamii} the authors considered the region 74--130 MeV.
Therefore, they could detect only a small part of the SND contribution.
Moreover, they used a very large
acceptance of the spectrometer which detected these protons, while the
protons under consideration have to fly in a very narrow angle cone.
As a result,
the ratio of the effect to the background in this work is more than 10
times worse than in \cite{prc,epj}.
Very big errors and absence of a proper cut on the energy
of the protons from the decay of the $pX_1$ state in
\cite{ tamii} did not allow the the authors to observe any structure in the
$pX_1$ mass spectrum.

It is worth noting that the reaction $pd\to NX$ was investigated in
other works too (see for example \cite{set}). However, in contrast to
\cite{prc,epj}, the authors of these works did not study
either the correlation between the parameters of the scattered proton and
the second
detected particle or the emission of the photon from the dibaryon decay.
Therefore, in these works the relative contribution of the dibaryons under
consideration was very small, which hampered their observation.

Now let us consider the attempts to calculate masses of SNDs and
determine their quantum numbers.

In the framework of the MIT bag model, Mulders et al. \cite{muld}
calculated
the masses of different dibaryons, in particular, $NN$-decoupled dibaryons.
They predicted dibaryons $D(T=0;J^P =0^{-},1^{-},2^{-};M=2.11$ GeV) \ and
$D(1;1^{-};2.2$ GeV) corresponding to the forbidden states $^{13}P_J$
and $^{31}P_1$ in the $NN$ channel.
However, the dibaryon masses obtained exceed the pion production threshold.
Therefore, these dibaryons mainly decay into the $\pi NN$ channel.

Using the chiral
soliton model, Kopeliovich \cite{kop} predicted that the masses of
$D(T=1,J^P=1^+)$ and $D(0,2^+)$
dibaryons exceeded the two nucleon mass by 60 and 90 MeV, respectively.
These values are lower than the pion production threshold.

In the framework of the canonically quantized biskyrmion model
Krupnovnickas {\em et al.} \cite{riska} obtained an indication on the
existence of one dibaryon with J=T=0 and two dibaryons with J=T=1 with
masses smaller than $2m_N+m_{\pi}$.

Unfortunately, all results obtained  for the dibaryon masses are very model
dependent and cannot reproduce the experimental data.

\section{Exotic baryons}

As was shown above, in the missing mass $M_{X_1}$ spectra three peaks
at 966, 986, and 1003 MeV had been observed.
On the other hand, the peak at $M_{X_1}=1003\pm 2$ MeV corresponds to
the peak found in \cite{tat} which was attributed to an exotic
baryon state $N^*$. Such a baryon state was first suggested in \cite{azim}.

In \cite{tat} Tatischeff {\em et al.} investigated
the reaction $pp\to\pi^+ X$. This experiment was carried out using the
proton beam at the Saturn Synchrotron and SPES3 facility \cite{com}.
The measurements were performed at energies
of $T_p=1520$, 1805, and 2100 MeV and at six angles, for each energy, from
$0^{\circ}$ up to $17^{\circ}$. Three peaks with widths about 5--8 MeV
have been observed in the missing
mass spectra of this reaction at $M_X=1004$, 1044, and 1094 MeV with a
statistic significance between 17 and 2 standard deviations. Two of these
masses are below the sum of the nucleon and pion masses.

If exotic baryons with anomalously small masses really exist,
the peaks observed at 966, 986, and 1003 MeV in \cite{epj}
might be a manifestation of
such states. This is not in contradiction to the interpretation of the
peaks in the $M_{pX_1}$ mass spectra of \cite{epj} as SNDs, since in
principle the SNDs could decay into $NN^*$. In this case the SND decay
width could be equal to a few MeV.

The existence of such exotic states, if
proved to be true, will fundamentally change our understanding of the
quark structure of hadrons.

Exotic baryon states with masses smaller than $m_N+m_{\pi}$ can mainly
decay with an emission of photons. If they decay into $\gamma N$ then
such states have
to contribute to the Compton scattering on the nucleon. However, L'vov and
Workman \cite{lvov} showed that existing experimental data on this process
"completely exclude" such exotic baryons as intermediate states in the
Compton scattering on the nucleon. On the other hand, the early Compton
scattering data were not accurate enough to rule out these baryon
resonances. Moreover, a measurement of the process $\gamma p\to\gamma p$
in the photon energy range
$60< E_{\gamma}<160$ MeV resulted in a peak at $M\approx 1048$ MeV
with an experimental resolution of 5 MeV and with 3.5 standard deviations
\cite{beck}. Unfortunately, the accuracy of this experiment is not enough
to do an unambiguous conclusion about the $N^*$ contribution to the Compton
scattering on the nucleon.

In \cite{kob} it was assumed that these states could belong to the
totally antisymmetric $\underline{20}$-plet of the spin-flavor SU(6)$_{FS}$
symmetry. Such a $N^*$ can transit into a nucleon only if two quarks from
the $N^*$ participate in the interaction \cite{feyn}
Then the simplest decay of the
exotic baryons with the small masses is $N^*\to \gamma\gamma N$.
The production of such baryons in reactions on the nucleon is possible only
through states with the mixed spin-flavor symmetry. The lowest of them are
$D_{13}(1520)$ and $S_{11}(1535)$ resonances.

The reaction $ep\to e^{\prime}\pi^+X^0$ was studied in \cite{jiang} to
search for narrow baryon resonances.
In this experiment the mass resolution of 2.0 MeV
was achieved. A search for structures in the mass region of
$0.97<M_{X^0}<1.06$ GeV yielded no  signal. This experiment was performed
at the invariant mass of the $\pi^+X^0$ system
equal to $W\approx 1.44$ GeV. However, if the excited baryons belong to
the totally antisymmetric representation, this value of $W$ is far enough
from the position of the $D_{13}(1520)$ and $S_{11}(1535)$ resonances
and the cross section of the $N^*$ production could be small at this $W$.

In \cite{kohl} the  $ep\to e^{\prime}\pi^+X$ and
$ed\to e^{\prime} pX$ reactions were
investigated at MAMI. The missing mass resolution
was 0.6 to 1.6 MeV (FWHM) in the proton experiment and 0.9 to 1.3 MeV
in the deuteron experiment. None of these measurements showed a signal for
narrow resonances to a level of down to $10^{-4}$ with respect to the
neutron peak in the missing mass spectra. Unfortunately, the value of
$W$ used in this experiment does not also allow the hypothesis of
\cite{kob} to be checked.

It should be noted that the reaction
$ed\to e^{\prime} pX$ is the analog of the process
$pd\to pD\to p \gamma d$ studied in
\cite{epj}. It has been shown in this work that the SND decay channel
$D\to \gamma d$ is strongly suppressed. Therefore, the negative result
of work \cite{kohl} at the investigation of the reaction
$ed\to e^{\prime} pX$ only supports
the result of \cite{epj} and it is an additional argument that the SNDs
observed in \cite{prc,epj} are isovectors.

On the other hand, the $N^*$s were produced in \cite{epj,tat}, more
probably, from the decay of 6-quark states,
what is supported by the observation
of the dibaryon resonances in \cite{epj}. Therefore, an exotic quark
structure of the $N^*$ could arise which suppressed, in particular,
the decay $N^*\to \gamma N$ and could be the reason that such states
were not observed in reactions on separate nucleons up to now.

In order to clarify the question about an existence of such exotic baryons,
different experiments were proposed, in particular, in \cite{mainz}.

In \cite{tat} it was shown that values of the masses of the
baryon resonances, observed in this work, can be reproduced with good
enough accuracy by the mass formula
for two colored clusters of quarks at the end of a stretched bag which was
derived in terms of color magnetic interactions \cite{muld,besl}.
There are two free parameters in this model which were fixed by requiring
the mass of the nucleon and that of the Roper resonance to be reproduced
exactly. As a result of the calculations, the following values of the
masses and possible isospin $(I)$ and spin $(J)$ of these
baryons have been obtained:
$$
 M(I;J)=1005(1/2;1/2,3/2),\qquad 1039(1/2;3/2).
$$

N. Konno \cite{kon} pointed out that the masses of the exotic baryons from
\cite{tat} can also be calculated with the help of the mass formula of
the diquark cluster model \cite{ueh}. Eight free parameters of this model
were fixed using data of baryon masses and the $\pi d$ phase shift.
This model predicted the following values of the masses and $I,J^P$:
$$
M=990(I=1/2(J^P=1/2^-); 3/2(1/2^-)),
$$
$$
M=1050(1/2(1/2^-,3/2^-); 3/2(1/2^-,3/2^-)),
\qquad M=1060(1/2(1/2^-,3/2^-)).
$$

However, these two models do not reproduce the values of masses: 966 and
986 MeV, obtained in \cite{epj}.

Th. Walcher \cite{walch} noted that the masses taking all experiments
together and
including the neutron ground state and two additional masses at 1023 and
1069 MeV are equidistant within the errors with an average mass difference
of $\Delta M=21.2\pm 2.6$ MeV. The author hypothesized the existence of a
light Goldstone boson with the mass of 21 MeV consisting of light current
quarks. It was assumed that the series of excited states is
due to the nucleon in its ground state plus 1, 2, 3,... light Goldstone
bosons as the quantum of excitation.

\section{ Mass formula for exotic baryons}

\begin{figure}
\epsfxsize=7cm       
\epsfysize=3cm          
\centerline{
\epsffile{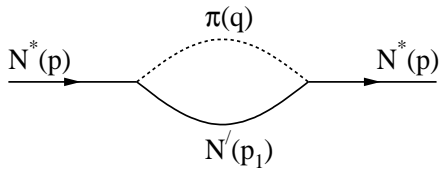}}
\caption{Baryon-meson contribution to $\Sigma(p)$.}
\label{loop}
\end{figure}

In this section we construct a model which allows the calculation of
the masses of all possible exotic baryon states below the $\pi$ meson
production threshold. This model is based on the
calculation of the contribution of a meson--baryon loop to the exotic
baryon mass operator \mbox{(fig. \ref{loop})} with the help of
dispersion relations with two subtractions.

An analysis of the mass shifts of experimentally well-known baryons due to
meson-baryon loops was carried out in a set of works (see for references
\cite{cap}). In these works, the self energy of a baryon was
calculated, as a rule, in the framework of a perturbation
theory. In this case, an underintegral expression diverges strongly and
additional assumptions about a behavior of baryon-baryon-meson vertices
are required.

We determine the mass operator as
\begin{equation}
G^{-1}(p)=\hat p-m-\Sigma(p), \qquad \Sigma(p)=a\hat p+b
\end{equation}
where $p$ is the 4-momentum of the $N^*$ under consideration, $m$
is the mass of the baryon in the intermediate state. Then
the mass of the $N^*$ is equal to
\begin{equation}
\label{M}
M=m+Re\,\delta(M)
\end{equation}
where
\begin{equation}
\delta(M)=\bar u(p)\Sigma(p)u(p)=\bar u(p)(aM+b)u(p)=aM+b,
\end{equation}
$a$ and $b$ are scalar functions.

In order to find the mass $M$, we construct the dispersion relations over
$M^2$ for $\delta(M)$ with two subtraction at $M^2=m^2$. Then taking into
account (\ref{M}) we obtain the following nonlinear integral equation for
$M$
\begin{eqnarray}
\label{ds}
M&=&m+Re\,\delta(m)+\left.(M^2-m^2)\frac{d\,Re\,\delta(M)}{d\, M^2}
\right|_{M=m}+ \nonumber \\
&&\frac{(M^2-m^2)^2}{\pi}P\!\!\!\int\limits_{(m+\mu)^2}^{\infty}
\frac{Im\,\delta(x)\,dx}{(x-M^2)(x-m^2)^2}.
\end{eqnarray}
where $P$ means a principal part of the integral.

All particles in the intermediate
state are on their mass shells.
As the subtraction is performed at the mass shell of the baryon in
the intermediate state,
the subtraction constant $Re\,\delta(m)$ is equal to zero. We also
assume that $dRe\,\delta(M)/dM^2|_m=0$. This assumption corresponds to
a supposition that the baryon with the mass $m$ is in the ground state.
It should be noted that if one takes
into account a few different baryons in the intermediate state, the
subtraction constants could not be equal to zero because
in this case the mass $m$ does not
coincide with the mass shell for some of these baryons.

Two subtractions in the dispersion relations provide a very good convergence
of the underintegral expression in (\ref{ds}). Therefore, we restrict
ourselves to a consideration of one baryon and the pion only in the
intermediate state. The calculations showed that the
contribution of the $\sigma$ meson is negligible. Therefore, it is expected
that the contribution of other heavy mesons
in the mass region under consideration is negligible too. However,
these contributions could be important in the mass region higher
than the meson--baryon production threshold.

Equation (\ref{ds}) can be also used for the calculation of the masses of
the well-known nucleon resonances. Therefore,
to check an efficiency of the mass formula obtained, we will apply it to
calculate the mass of $\Delta(1232)$ resonance.

The contribution of the pion--nucleon loop to $Im\,\delta_{\Delta}(M)$ can
be written as
\begin{equation}\label{imd1}
\bar u_{\mu}(p)Im\,\delta_{\Delta}(M)u_{\mu}(p)=\frac12\left(
\frac{g^2_{\Delta}(M)}{4\pi}\right)\frac{1}{\mu^2}\bar u_{\mu}(p)q_{\mu}(
\hat p_1+m)q_{\lambda}u_{\lambda}(p)
\end{equation}
where $p_1$ and $m$ ($q$ and $\mu$) are the 4-momentum and the mass of
the nucleon (pion) in the intermediate state, $p$ is the 4-momentum of
the $\Delta(1232)$ resonance.

Multiplying this expression by $u_{\nu}(p)$ from the left and by
$\bar u_{\nu}(p)$ from the right and using the condition
$u_{\nu}(p)\bar u_{\mu}(p)=\Delta_{\nu\mu}(p)$ we have
\begin{equation}\label{imd2}
\Delta_{\nu\mu}(p)\,Im\,\delta_{\Delta}(M)\,\Delta_{\mu\nu}(p)=\frac12
\frac{g^2_{\Delta}(M)}{4\pi}\frac{1}{\mu^2}\Delta_{\nu\mu}(p)q_{\mu}
(\hat p_1+m)q_{\lambda}\Delta_{\lambda\nu}(p)
\end{equation}
where
$$
\Delta_{\nu\mu}(p)=\frac{1}{3M^2}(\hat p+M)\left[2p_{\nu}p_{\mu}-
3M^2g_{\nu\mu}+M^2\gamma_{\nu}\gamma_{\mu}+M(\gamma_{\nu}p_{\mu}-
\gamma_{\mu}p_{\nu})\right].
$$
Calculating traces in the left and right parts of (\ref{imd2})
and taking into account that $p^2=M^2$ we obtain
\begin{equation}\label{imd3}
Im\,\delta(M)=-\frac16\left(\frac{g^2_{\Delta}(M)}{4\pi}\right)\frac{1}
{\mu^2}|q|^2(m+E_1),
\end{equation}
here $|q|=\sqrt{E_1^2-m^2}$ .

In accordance with \cite{blom,bar} we use the following
parametrization
\begin{equation}\label{par}
g^2_{\Delta}(M)=g^2_0\frac{1+R^2|q_{\Delta}|^2}{1+R^2|q|^2},
\end{equation}
$$
|q_{\Delta}|=|q(M=M_{\Delta})|,\qquad M_{\Delta}=1232\: {\rm MeV},
\qquad R=5.5\: {\rm GeV}^{-1}.
$$
The coupling constant $g^2_0$ can be expressed through the decay width
of the $\Delta^+(1232)$ resonance $\Gamma_{\Delta}$ as
$$
\frac{g^2_0}{4\pi}\frac{1}{\mu^2}=
\frac{9M_{\Delta}\Gamma_{\Delta}}{|q_{\Delta}|^3(E_{\Delta}+m)}
$$
where $E_{\Delta}=\sqrt{|q_{\Delta}|^2+M^2_{\Delta}}$. This value of
$g^2_0/(4\pi)\mu^{-2}$ is equal to $1.178\,\mu^{-2}$ for the decay channel
$\Delta^+(1232)\to \pi^+ n$ and $1.055\,\mu^{-2}$ for the
$\Delta^+(1232)\to \pi^0p$ channel.

Substituting (\ref{imd3}) into (\ref{ds}) and taking into
account the indicated parametrization we obtain for the mass of
the $\Delta$ resonance the value of 1247 MeV. It differs not strongly
from the experimental value of 1232 MeV. This result confirms a good
efficiency of this mass formula.

\section{Calculation of the masses of the exotic baryons}\label{exbar}

We will consider here only baryons with the spin equal to 1/2.
Then the function $Im\,\delta$ for the baryon-pion loop can be written as
\begin{equation}  
\bar u(p)Im\,\delta(M)u(p)=\bar u(p)Im\,\Sigma(M)u(p)=
N\bar u(p)(\mp\hat p_1+m)u(p)
\end{equation} 
where $N=1/2(g^2/4\pi)|p_1|/M$,
$p_1$ is the 4-momentum of the baryon in the intermediate state.
The sign minus (plus) at $\hat p_1$
corresponds to the same (opposite) parities of the final baryon and the
baryon in the intermediate state.
Multiplying this expression by $u(p)$ from the left and by
$\bar u(p)$ from right and using the condition
$u(p)\bar u(p)=(\hat p+M)/2M$ we have
\begin{equation}
(\hat p+M)Im\,\delta(M) (\hat p+M)=N(\hat p+M)(\mp\hat p_1+m)(\hat p+M).
\end{equation}
Calculating traces in the left and the right parts of this expression
we obtain:
\begin{equation}
Im\,\delta(M)=\frac{N}{2M^2}[\mp 2(pp_1)M+(p^2+M^2)m].
\end{equation}
As $p^2=M^2$ , $2(pp_1)=M^2+m^2-\mu^2=2M E$, we have
\begin{equation}\label{imb}
Im\,\delta(M)=\frac12\frac{g^2}{4\pi}\frac{|p_1|}{M}(m\mp E),
\end{equation}
here $|p_1|=\sqrt{E^2-m^2}$.

In our calculations we
assumed that all baryons under consideration have the isotopic
spin and the spin equal to $1/2$. Then equation (\ref{ds}) has solutions
in the mass region lower than the $\pi N$ production threshold only if
the final baryon with the mass $M$ and the baryon in the intermediate
state have opposite parities.

Taking the nucleon plus the pion ($\pi^+$ and $\pi^0$) in the
intermediate state, we find for the first exotic baryon state
the mass $M=963.4$ MeV and
$J^P=1/2^-$. Then taking the first exotic baryon with $m=963.4$ MeV,
$J^P=1/2^-$ and the $\pi$ meson in the intermediate state
we obtain for the second exotic baryon state $M=987.0$ MeV and
$J^P=1/2^+$.
Continuing the same procedure, six possible exotic states of baryons
have been found. The states obtained are listed in table 1 where the
experimental data are also given.
\begin{table}
\centering
\caption{\label{table1}
The masses and $J^P$ of the exotic baryon resonances.}
\begin{tabular}{|c|c|c|c|c|}\hline
     &          &  model   &  experiment & experimental \\
$N^*$& $J^P$    &$M$ (MeV) & $M$ (MeV)   & works        \\ \hline
 1   &$\frac12^-$& 963.4   & $966\pm 2$  & \cite{epj} \\ \hline
 2   &$\frac12^+$& 987   & $986\pm 2$  & \cite{epj} \\ \hline
 3   &$\frac12^-$& 1010  & $1004$ &\cite{tat,epj}\\ \hline
 4   &$\frac12^+$& 1033    &    ?        &               \\ \hline
 5   &$\frac12^-$& 1056    & $1044$ & \cite{tat}    \\ \hline
 6   &$\frac12^+$& 1079    &    ?        &               \\ \hline
\end{tabular}
\end{table}

As seen from this table, the results of the calculations are in a good
agreement with the experimental data. The mass values of two still
unobserved states at 1033 and 1079 MeV are close to the ones predicted in
\cite{walch}.

At the calculation we assumed that the coupling constant
$g^2/4\pi$ in the vertices $NN^*\pi$ and $N^*_iN^*_f\pi$
is same for all $N^*$ and
equal to the coupling constant of the $\pi NN$ interaction
($g^2_{pp\pi^0}/4\pi=14.6$). So, the increase of the difference between
the model predictions for the masses and experimental data
with a rise of the mass could be caused, in
particular, by this assumption. Therefore, it is expected that the real
value of the mass for the still unobserved state \#6 should be smaller by
$\sim 10$ MeV and so it has to be bellow the pion production threshold.

As follows from the table, the odd parity has been predicted for
all baryon states found in \cite{tat}.
It agrees with the comment of Th. Walcher \cite{walch}
and is due to the kinematics of the experiment \cite{tat}. This experiment
detected $\pi^+$ and $p$ from the reaction $pp\to\pi^++pX$ in coincidence
in one spectrometer at small angles. Since the $N^*$ states
were mainly observed in the forward direction with respect to the beam
($\theta\approx 5^{\circ}$), this means that the outgoing $\pi$ and $p$
carry no angular momentum and only the odd parity state of $N^*$ could be
observed in this experiment. But the baryons with the odd parity
cannot belong to the totally antisymmetric $\underline{20}$-plet
\cite{kob}.

It should be noted that the big value of the constant $g^2/4\pi$
could suppose the noticeable contribution of the $N^*$ to the elastic
$\pi N$ scattering. However, the experimental data on this process
do not support such a possibility \cite{strak}. Therefore, if the model
is correct, there are two possibilities. The first,
the constant $g^2/4\pi$ is the effective coupling constant of some
complicated
interaction of a pion cloud with the baryon in the intermediate state
and does not connect with the usual pion--nucleon coupling constant.
The second, the exotic baryons with the anomaly small masses do not
exist. In this case, equation (\ref{ds}) describes the effective masses
corresponding to the positions of the narrow peaks in the phase space of
the products of the SNDs decay. Such peaks are not real resonances and
do not contribute to processes on separate nucleons
and the value of the constant $g^2/4\pi$
coincides with the coupling constant of the $\pi NN$ interactions by
accident.

\section{Mass formula for the SNDs}

Using the complete Green function of the dibaryons
$$
\Delta(p^2)=\frac{F(p)}{p^2-m^2-\delta_D(p^2)},
$$
we determine the SND mass as
\begin{equation}\label{mdb}
M^2=m^2+Re\,\delta_D(M^2)
\end{equation}
where $\delta_D(M^2)$ is the self energy of the SND under study and
$m$ is mass of the dibaryon in the intermediate state.

The self energy of the lightest SND will be determined in one loop
approximation through the interaction of the pion with the  6-quark
state of the deuteron.
The self energy of the next SND will be obtain through the interaction of
the pion with the lightest SND and so on.

We will calculate the SND self energy with the help of
the dispersion relations with two subtractions at $M^2=m^2$.
Then taking into account (\ref{mdb}) we obtain the mass formula for the
SNDs
\begin{eqnarray}
\label{dsd}
M^2&=&m^2+Re\,\delta_D(m^2)+\left.(M^2-m^2)\frac{
d\,Re\,\delta_D(M^2)}{d\,M^2}\right|_{M^2=m^2}+ \nonumber \\
&&\frac{(M^2-m^2)^2}{\pi}P\!\!\!\int\limits_{(m+\mu)^2}^{\infty}
\frac{Im\,\delta_D(x)\,dx}{(x-M^2)(x-m^2)^2}.
\end{eqnarray}
As the subtraction is performed at the mass shell of the
dibaryon in the intermediate state, the subtraction function
$Re\,\delta_D(m^2)$ is equal to zero. Assuming that this dibaryon is
in the ground state, we have
$\left.d\,Re\,\delta_D(M^2)/d\,M^2\right|_{M^2=m^2}=0$.

There are two possibilities to calculate $Im\,\delta_D(x)$.
The first, when the SND under study and the dibaryon in the intermediate
state have opposite parities. The second, when these parities are the same.

The vertices $D^{\prime}(1^{\mp})\to\pi+D(1^{\pm})$  and
$D^{\prime}(1^{\pm})\to\pi+D(1^{\pm})$ can be respectively written as
\begin{equation}
\Gamma^{(-)}=\frac{g_1}{M}G_{\mu\nu}\Phi^{\mu\nu}, \qquad
\Gamma^{(+)}=\frac{g_2}{M}\epsilon_{\mu\nu\lambda\sigma}
G^{\mu\nu}\Phi^{\lambda\sigma}
\end{equation}
where $G_{\mu\nu}=v_{\mu}p_{\nu}-p_{\mu}v_{\nu}$,
$\Phi_{\mu\nu}=w_{\mu}p_{1\nu}-p_{1\mu}w_{\nu}$; $v$ and $w$ are the
4-vectors of the polarization of the final SND and the dibaryon in
the intermediate states, respectively; $p$ and $p_1$ are their 4-momenta.

As result of the calculations we have the following expression for
the imaginary part of $\delta_D(x)$ when the final SND and dibaryon
in the intermediate state have the opposite parities
\begin{equation}\label{oppos}
Im\,\delta_D^{(-)}(x)=\frac13\left(\frac{g^2_1}{4\pi}\right)
\frac{q[(x+m^2-\mu^2)^2+2m^2x]}{x^{\frac32}}
\end{equation}
where $q$ is the pion momentum equal to
$q=[(x-(m+\mu)^2)(x-(m-\mu)^2]^{1/2}/2x^{1/2}$.

For the dibaryons with the same parities we have
\begin{equation}\label{same}
Im\,\delta_D^{(+)}(x)=\frac83\left(\frac{g^2_2}{4\pi}\right)
\frac{q[(x+m^2-\mu^2)^2-4m^2x]}{x^{\frac32}}
\end{equation}

The coupling constant $g^2_1/4\pi$ in the vertex for transition of
the 6-quark state of the deuteron $(D(0,1^+))$ plus the pion to the
SND $D(1,1^-)$ has been fixed by requiring a reproduction of the mass
$M=1904$ MeV. It results in
\begin{equation}\label{const1}
\frac{g^2_1}{4\pi}=26.585
\end{equation}

As follows from estimations of the value of the cross section of
the $D(1,1^-,M=1904)$ production in the process $pd\to p+pX_1$
\cite{epj,tamii}, the product $(g^2_1/4\pi\;\eta)$ is equal to
$\sim 2\times 10^{-3}$ where $\eta$ is the probability of the 6-quark
state existence in the deuteron. Then taking into account the value
(\ref{const1}) for the coupling constant $g^2_1/4\pi$, one obtains
$\eta\sim 1\times 10^{-4}$. This value of $\eta$ is essentially
smaller than its estimation found in \cite{kondr} ($\eta\le 0.03$).
As a result, the decay width of the SNDs with small masses might be
considerably smaller than 1 eV.

We will assume that the all SNDs under consideration have the isospin
$T=1$ and $J^P=1^{\pm}$. The coupling constants $\bar g_1$ in the vertex
$D(1,1^{\pm})\to D(1,1^{\mp})+\pi$ differs from $g_1$.

On the other hand, now two channels give the contribution to the
imaginary part $Im\,\delta_D(x)$. For example, such
a contribution to the self energy of a positive charged supernarrow
dibaryon $D^+$ is given by the channels
$$
D^0+\pi^+ \quad {\rm and} \quad D^++\pi^0.
$$
The calculations in the frame of our model give the very close values of
the masses of $D^0$ and $D^+$ dibaryons. Therefore, we will take them 
equal one to other.

In order to calculate the mass of the following SND, we take in the
intermediate state the SND $D(1,1^-)$ with $M=1904$ MeV and the pion.
If the constant $\bar g^2_1/4\pi=3/4 (g^2_1/4\pi)$ then taken into account
the contribution from two possible channels and using (\ref{oppos})
we find $M=1924$ MeV and $J^P=1^+$. To calculate the mass of the third SND
we consider in the intermediate state the SND $D(1,1^+)$ with $M=1924$ MeV
and the pion ($\pi^+$ and $\pi^0$) and use the same value of
$\bar g^2_1/4\pi$ and (\ref{oppos}). It results in $M=1943$ and $J^P=1^-$.
Continuing such calculations with
(\ref{oppos}) and using the same value of $\bar g_1$ we get other
values of the SND masses. The results of the calculations of the masses
and $J^P$ of the SNDs are listed in table \ref{dibm}.
\begin{table}
\centering
\caption{The masses and $J^P$ of the SNDs.}
\label{dibm}
\begin{tabular}{|c|c|c|c|c|}\hline
     &     &  model  &  experiment & experimental \\
$No$& $J^P$&$M$ (MeV)& $M$ (MeV)   &  works       \\ \hline
 1   &$1^-$& 1904    &$1904\pm 2$  & \cite{epj}   \\ \hline
 2   &$1^+$& 1924    &$1926\pm 2$  & \cite{epj}   \\ \hline
 3   &$1^-$& 1943    &$1942\pm 2$  & \cite{epj}   \\ \hline
 4   &$1^+$& 1962    &$1956\pm 6$  & \cite{khr}   \\ \hline
 5   &$1^-$& 1982    & $1982$ &predicted \cite{epj,tat2}\\ \hline
 6   &$1^+$& 2001    &    ?        &               \\ \hline
\end{tabular}
\end{table}

As seen from this table, the values of the SND masses obtained are in
a good agreement with the experimental data. The existence of the SND with
the mass $M=1982$ MeV has been predicted in \cite{epj,tat2}
as a consequence of the observation of the peak in the missing mass spectra
of the reaction $pp\to\pi^+pX$ \cite{tat} at $M_X=1044$ MeV.

Our calculations predict also the possibility of the existence of
the new SND $D(1,1^+)$ with the mass $M=2001$ MeV.

As the SNDs observed in \cite{prc,epj} decay into $NN^*$ with the small
relative momentum between $N$ and $N^*$, the SND parity has to be
determined by the parity of the $N^*$. As seen from tables 1 and
\ref{dibm}, the SND parities found here agree with the results obtained for
the $N^*$ states.

As for the calculations with $Im\,\delta^+_D(x)$, the results very close
to the ones presented in \mbox{table \ref{dibm}} can be  obtained if
$g_2^2/4\pi=3g_1^2/4\pi$ and $\bar g_2^2/4\pi=3\bar g_1^2/4\pi$.

Then the calculation of the mass of the SND due to the pion-deuteron loop
has given $D(1,1^+)$ with $M=1925$ MeV. To calculate the mass of the
next SND, we take $D(1,1^-)$ with $M=1904$ and
the $\pi^+$ and $\pi^0$ mesons in the intermediate state. It results in
$D(1,1^-)$ with the mass $M=1939$ MeV. Then taking from
table \ref{dibm} the following SND $D(1,1^+)$ with $M=1924$ MeV and
the pions in the intermediate state, we obtain $D(1,1^+)$ with $M=1958$
MeV. Continuing this procedure we find also $D(1,1^-)$ with $M=1978$ MeV
and $D(1,1^+)$ with $M=1998$ MeV. These values of the masses obtained with
the help of (\ref{same}) are very close to the ones calculated using
(\ref{oppos}) for the imaginary part of $\delta_D(x)$
(see table \ref{dibm}).
However, it is not possible to obtain SND $D(1,1^-)$ with $M=1904$ by
using eq. (\ref{same}) only.

\section{Conclusion}

As a result of the study of the reaction $pd\to p+pX_1$,  three narrow
peaks at $M_{pX_1}=1904$, 1926, and 1942 MeV have been observed 
\cite{prc,epj}. The analysis
of the angular distributions of the protons from the decay of the $pX_1$
states showed that the peaks found can be explained as a manifestation of 
the isovector SNDs, the decay of which into two nucleons is forbidden by 
the Pauli exclusion principle. The observation of the peaks in the 
missing mass $M_{X_1}$ spectra at 966, 986, and 1003 MeV is an additional 
indication that the dibaryons found are the SNDs.

On the other hand, these peaks in $M_{X_1}$ mass spectra and peaks observed
in \cite{tat} in the reaction $pp\to\pi^+pX$
could be consider as the new exotic baryon states with small
masses. However, additional experiments are necessary to understand the
real nature of these peaks.

In the present paper, the mass formulae for the SNDs and the exotic baryons
have been constructed using the dispersion relations with two subtraction
for their self energies in the one loop approximation. These mass
formulae were used to calculate the masses and determine parities
of the SNDs and the exotic baryons.
The obtained values of the masses are in a good agreement with the
experimental data of \cite{epj,tat,khr}.
The application of the mass formula found to calculate mass of the
$\Delta(1232)$ resonance has given the value close enough to the
experimental one.

\section*{Acknowledgement}

The author thanks R. Beck, V. Fainberg, V. Fetisov, A. Kobushkin,
B. Tatischeff, and Th. Walcher for helpful discussions.
This work was supported by RFBR, grant No. 03-02-04018

\end{document}